\documentclass[fleqn,12pt]{article}

\usepackage{cite}
\usepackage{epsfig}
\usepackage{a4wide}

\def\MSbar{\overline{\mathrm{MS}}}
\def\as{\alpha_s}
\def\ep{\epsilon}

\def\nl{{n^{}_{\! l}}}
\def\nh{{n^{}_{\! h}}}

\begin{document}
  
    \title{\bf Tops from Light Quarks: \\
      Full Mass Dependence at Two-Loops in QCD}

    \author{M. Czakon \\ [.5cm]
      Institut f\"ur Theoretische Physik
      und Astrophysik, Universit\"at W\"urzburg, \\
      Am Hubland, D-97074 W\"urzburg, Germany \\ [.5cm]
      Institute of Physics, University of Silesia, \\
      Uniwersytecka 4, PL-40007 Katowice, Poland}

    \maketitle

    \begin{abstract}

      I present the two-loop QCD corrections to the production of a
      massive quark-anti-quark pair in the massless quark-anti-quark
      annihilation channel. The result is obtained as a combination of a deep
      expansion in the mass around the high energy limit and of a numerical
      integration of a system of differential equations. The primary
      application of the outcome and developed methods is top quark pair
      production at the Large Hadron Collider.

    \end{abstract}
    
\section{Introduction}

One of the most important goals of the Large Hadron Collider (LHC) is the
measurement of top quark properties. This will be possible thanks to the
abundant statistical samples reaching about 8 million pairs produced per year
in the low luminosity phase. Besides the mass and decay parameters, the total
cross section constitutes a primary observable. The experimental prospects of
obtaining a measurement accuracy below ten percent for this quantity put a
high demand on theoretical predictions of matching quality. At present, the
known next-to-leading order corrections \cite{Nason:1987xz} have an error
estimated from scale variation at about 12\%. Soft gluon resummation
\cite{Laenen:1991af,Bonciani:1998vc,Kidonakis:2003qe}, which has been an
excellent tool for the Tevatron, and helped reduce the error to about 5\%, is
not safely applicable in the framework of the LHC. This is due on the one hand
to the higher energy and on the other to the dominance of the gluon flux over
the quark flux. Furthermore, the mentioned  high statistics warrant the
preparation of a Monte-Carlo generator of suitable precision, which cannot
rely on resummed cross sections.

In view of these facts it is necessary to provide a result for the
next-to-next-to-leading order cross section, at best in a fully differential
form. This requires four separate parts at the partonic level: the two-loop
virtual corrections, the one-loop squared virtual corrections, the one-loop
real-virtual corrections with an additional parton in the final state, and the
tree-level real corrections with two additional partons in the final
state. Out of these, the second part is known from
\cite{Korner:2005rg,Korner:2008bn}, the third from \cite{Dittmaier:2007wz},
where the next-to-leading order corrections to the $t\bar{t}$+jet corrections
have been derived. Unfortunately, as part of a cross section calculation for
top quark pair production this result still needs subtraction terms in order
to allow for integration over the full phase space. Finally, while there is no
result for the real radiation, the two-loop corrections\footnote{The case of
  massless quark production has been studied in a number of papers
\cite{Anastasiou:2000kg,Anastasiou:2000ue,Anastasiou:2001sv,
Glover:2004si,DeFreitas:2004tk,Bern:2003ck}} have been recently evaluated in
the limit of small top quark mass \cite{Czakon:2007ej,Czakon:2007wk}. This
result is applicable for highly energetic tops, for example when high $p_T$
cuts would be applied. The bulk of events comes, however, from the region much
nearer to the partonic threshold.

In this paper, I present a complete result for the two-loop corrections in the
quark annihilation channel valid in the whole kinematically allowed region. It
has to be stressed that obtaining an amplitude expressed in analytic form
through some special functions seems out of reach in the nearest future. Since
the LHC will soon become operational it is necessary to resort to
semi-analytic/semi-numeric methods. The method adopted here is a combination
of a deep expansion in the mass around the high energy limit, which contains the
power corrections to the result of \cite{Czakon:2007ej}, and of numerical
integration of differential equations.

In the next section, I will first give a few definitions and then describe the
power corrections. A detailed study of the numerical methods and the full
result will follow in the last section of the main text.

\section{Power corrections}

The notation of this paper follows closely that of
\cite{Czakon:2007ej}. However, I reproduce all the necessary definitions for
the convenience of the reader.

The process under scrutiny is massless quark-anti-quark annihilation into a
massive quark-anti-quark pair
\begin{equation} q(p_1) + {\bar q}(p_2) ~ \rightarrow ~ Q(p_3,m) + {\bar
Q}(p_4,m) \, .
\end{equation}
The amplitude can be described with the help of Mandelstam variables
\begin{equation}
s = (p_1+p_2)^2 , ~ t  = (p_1-p_3)^2 - m^2 , ~ u  = (p_1-p_4)^2 - m^2 .
\end{equation}
Notice that the mass subtraction in the definition of the $t$ and $u$
variables was irrelevant for the results of \cite{Czakon:2007ej}, because only
logarithmic terms in the mass have been retained there. The advantage of this
definition lies in the symmetric range of variation
\begin{equation}
|t|, |u| \in \left[ \frac{s}{2}(1-\beta) , ~ \frac{s}{2}(1+\beta) \right] ,
\end{equation}
where $\beta$ is the velocity
\begin{equation}
\beta = \sqrt{1-\frac{4m^2}{s}} .
\end{equation}
The results will be parameterized by two dimensionless ratios
\begin{equation}
m_s = \frac{m^2}{s} , ~ x = -\frac{t}{s} .
\end{equation}
The additional scale introduced by dimensional regularization, $\mu$, has been
kept in the results as unspecified, but for the plots and numerical values
reproduced in the following $\mu = m$. The renormalization has been performed
in the $\MSbar$ scheme with $n_l$ massless and $n_h$ massive active
flavors. The necessary constants are known in the literature with sufficient
precision: the strong coupling renormalization to four-loop accuracy
\cite{vanRitbergen:1997va,Czakon:2004bu}, and the mass and field
renormalization of the  heavy quark in the on-shell scheme to three-loop
accuracy \cite{Chetyrkin:1999ys,Melnikov:2000qh,Melnikov:2000zc}. The
renormalization of the light quark field, which is non-vanishing because of
the presence of closed heavy quark loops has been explicitely given in
\cite{Czakon:2007ej} with two-loop accuracy.

After expanding the amplitude in the strong coupling constant up to the second
order
\begin{eqnarray}
  | {\cal M} \rangle
  & = &
  4 \pi \as \biggl[
  | {\cal M}^{(0)} \rangle
  + \biggl( {\as \over 2 \pi} \biggr) | {\cal M}^{(1)} \rangle
  + \biggl( {\as \over 2 \pi} \biggr)^2 | {\cal M}^{(2)} \rangle
  + {\cal O}(\as^3)
  \biggr],
\end{eqnarray}
the interesting, two-loop term, contracted with the Born amplitude can be
decomposed into color factors
\begin{eqnarray}
&& \!\!\!\!\!\!\!\!\!\!\!\! {\cal A}^{(0,2)} = 2 {\rm Re}\, \langle {\cal
  M}^{(0)} | {\cal M}^{(2)} \rangle = 2 (N^2-1) \\
&& \!\!\!\!\!\!\!\!\!\!\!\! \times \biggl(
N^2 A + B  + {1 \over N^2} C
+ N \nl D_l + N \nh D_h
+ {\nl \over N} E_l + {\nh \over N} E_h + \nl^2 F_l + \nl \nh F_{lh} + \nh^2 F_h
\biggr) .
\nonumber
\end{eqnarray}
The leading behavior of the amplitude in the limit $m \rightarrow 0$, has
been derived in \cite{Czakon:2007ej} using two different approaches. The first
is based on factorization properties of QCD amplitudes \cite{Mitov:2006xs},
and exploits a relation between the massless and massive cases. Unfortunately,
it does not give a handle on mass corrections or the full mass dependence. The
second approach is based on a direct evaluation of occurring integrals and is
an evolution of a strategy developed for Bhabha scattering
\cite{Czakon:2004wm,Czakon:2006pa}. The procedure starts with a reduction of
the integrals to masters with the help of the Laporta algorithm
\cite{Laporta:2001dd}, and subsequent expansion of the masters in the mass by
passing through Mellin-Barnes representations
\cite{Smirnov:1999gc,Tausk:1999vh}. The bulk of the work is performed by
Mathematica packages {\tt MBrepresentation} \cite{MBrepresentation} and {\tt
MB} \cite{Czakon:2005rk} together with further associated software.

The derivation of the asymptotic behavior of Mellin-Barnes representations is
performed recursively by closing the contours of integration and taking
residues. It is obvious that arbitrarily high orders of expansion in the mass
can be obtained. However, the coefficients will still be integrals requiring
evaluation by summation of infinite series or by some other
method. Previously, this last step has been completed with a combination of
{\tt XSummer} \cite{Moch:2005uc} and the {\tt PSLQ} algorithm \cite{PSLQ}. It
has to be stressed that every next order in $\ep$ contains more integrals with
a more complicated integrand structure. Fortunately, there exists an
alternative approach based on differential equations
\cite{Kotikov:1990kg,Remiddi:1997ny}.

Clearly, applying a derivative with respect to any invariant or the mass
introduces higher powers of denominators and/or numerator
structures. Furthermore, any set of integrals differing only by powers of
denominators and numerators can be reduced to a smaller set of masters. In
consequence one can write the following differential equation systems for the
coefficients of the Laurent expansion of the master integrals
\begin{eqnarray}
\label{eq:differential}
\frac{d}{d m_s} I_i(m_s, x) &=& \sum_j J^M_{ij}(m_s, x) I_j(m_s, x) \\
\frac{d}{d x} I_i(m_s, x) &=& \sum_j J^X_{ij}(m_s, x) I_j(m_s, x).
\end{eqnarray}
The Jacobian matrices, $J^{M}$ and $J^{X}$, have rational function
elements and it is implied that any master integral is a combination of the
$I_i(m_s,x)$ functions
\begin{equation}
M_i(m_s,x,\ep) = \sum_{j = k}^l \ep^j I_{i_j}(m_s,x).
\end{equation}
The lowest power of $\ep$ in the sum is fixed by the singularities of the
integral and cannot exceed $-4$ at this level of perturbation theory, whereas
the highest power is defined by the coefficient in the amplitude (there are spurious 
poles). In practice, the deepest expansion in $\ep$ that occurred was down to
order $\ep^3$ due to the particular choice of master integrals.

The differential equations Eq.~\ref{eq:differential} allow, in principle, to
fix the complete functional dependence of the master integrals. Unfortunately,
the functions present in the solution are not known at present, and therefore
a direct integration of the system has to be postponed. Nevertheless, the
differential equations in $m_s$ can be solved by means of a series expansion,
with boundaries given by the small mass limit as needed for the results of
\cite{Czakon:2007ej}. Following this idea, it is possible to derive
arbitrarily deep expansions of the amplitude in the mass, and thus provide the
power corrections, which were out of reach of the factorization approach. Of
course, the size of the intermediate expressions, combined with the available
computing resources puts a natural cut-off on the highest power that can be
computed. In fact, I have computed eleven terms of the series up to $m_s^{10}$.

\begin{figure}[t]
  \begin{center}
    \epsfig{file=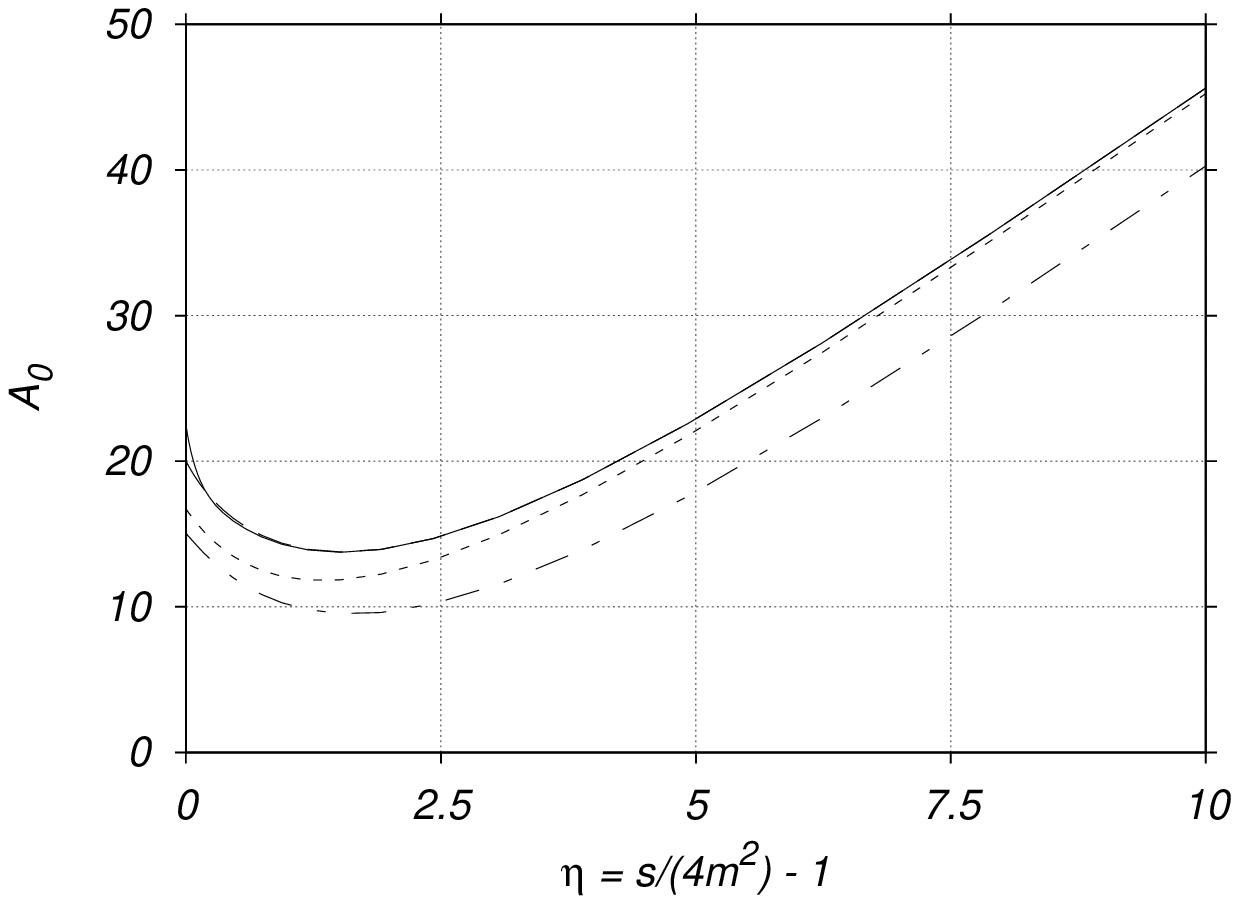,width=7cm} \hspace{ 1cm}
    \epsfig{file=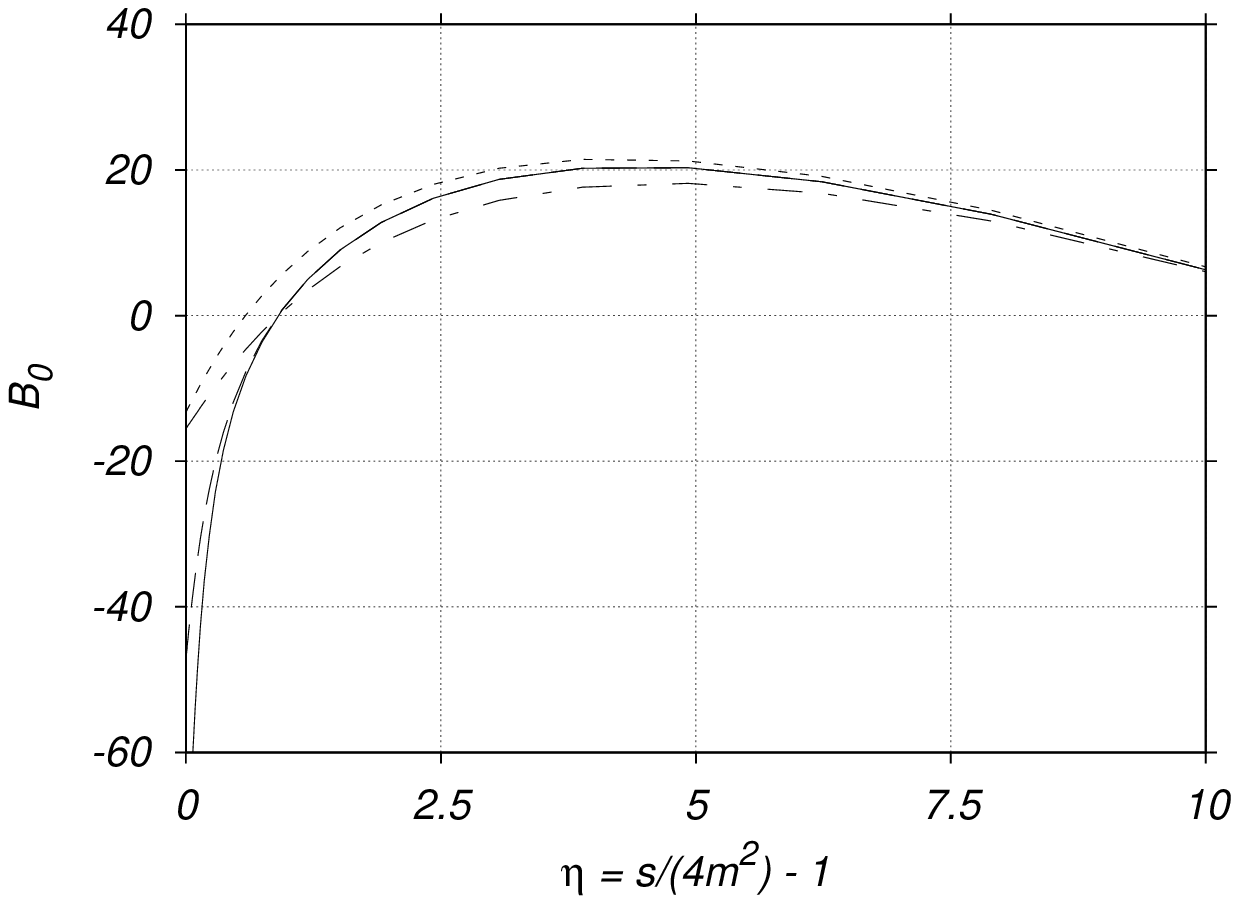,width=7cm} \\ \vspace{1cm}
    \epsfig{file=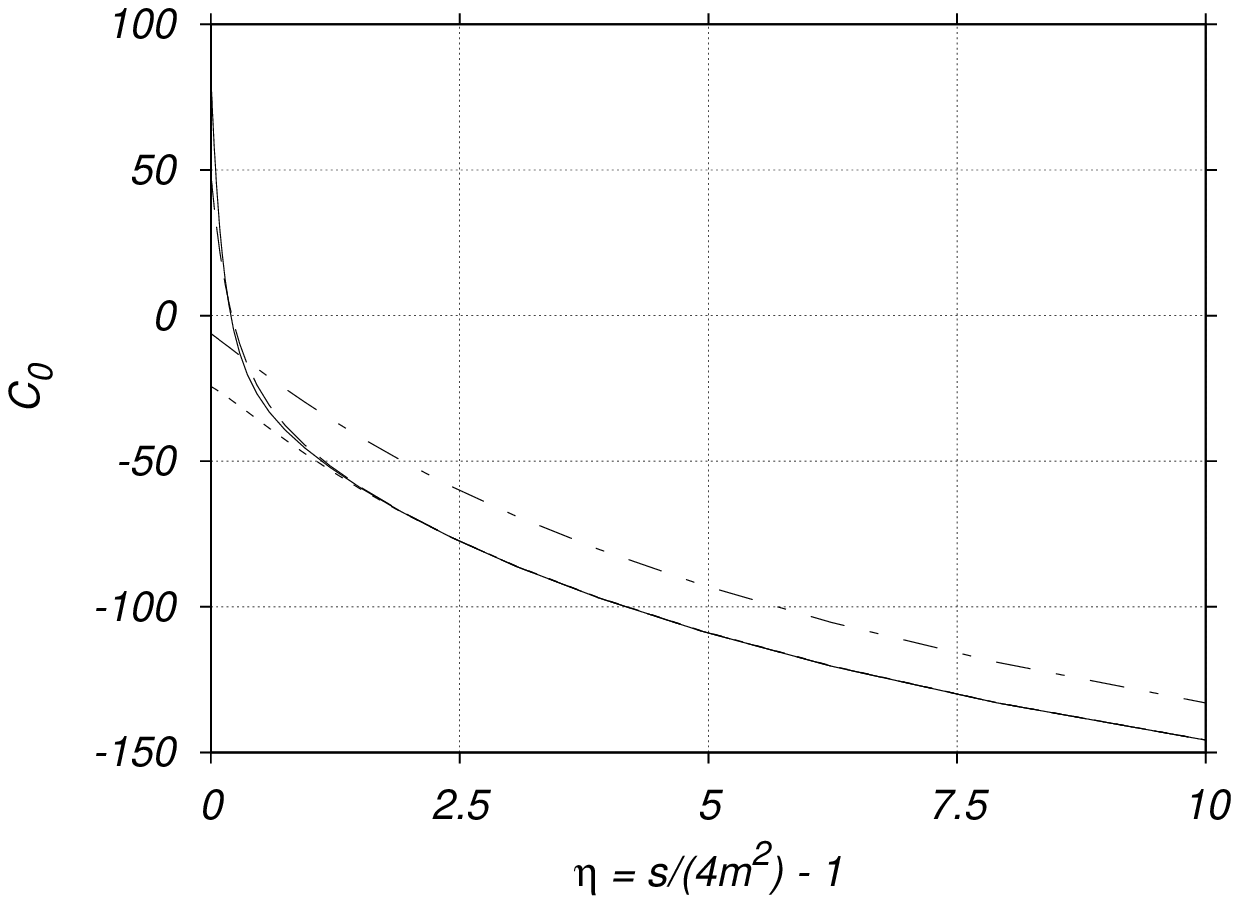,width=7cm}
  \end{center}
  \caption{\label{fig:expansion} Finite parts of the purely bosonic
    contributions  to the two-loop amplitude as expansions around the small mass
    limit at $x=1/2$. The solid line represents the eleven terms of the series as
    derived for the present publication, the long dashed - ten terms of the
    series, the short dashed - six terms of the series, the dash-dotted - the
    leading behavior.}
\end{figure}

The results for the finite parts (in the $\ep$ expansion) of the three purely
bosonic contributions $A$, $B$ and $C$, which are also the most involved as
far as the computation is concerned are shown in Fig.~\ref{fig:expansion}. The
plots correspond to 90 degree scattering, {\it i.e.} $x=1/2$. It is striking
that the series do not obviously diverge, which is usually the case with this
type of expansions. In fact, the series expansion for the leading color term,
$A$, is at worst asymptotic at threshold, and can still be used to obtain an
estimate accurate to a few percent at this point. On the other hand, the
growth of the subleading color coefficients is indicative of the true
behavior, but incorrect. As I will show in the next section, there is a true
divergence due to the Coulomb singularity, which cannot be reproduced with a
small mass expansion.

The small mass expansion is, in reality, not an expansion in $m^2/s$, but
rather in max($m^2/s$, $-m^2/t$, $-m^2/u$). For small $m$ and at the
kinematical boundary corresponding to forward scattering
\begin{equation}
  -\frac{m^2}{t} = -\frac{2m^2}{s\left(1-\sqrt{1-\frac{4m^2}{s}}\right)}
  \approx 1.
\end{equation}
A similar relation holds for $-m^2/u$ for backward scattering. In consequence,
the series will be asymptotic at best at the kinematic boundaries.

\begin{figure}[t]
  \begin{center}
    \epsfig{file=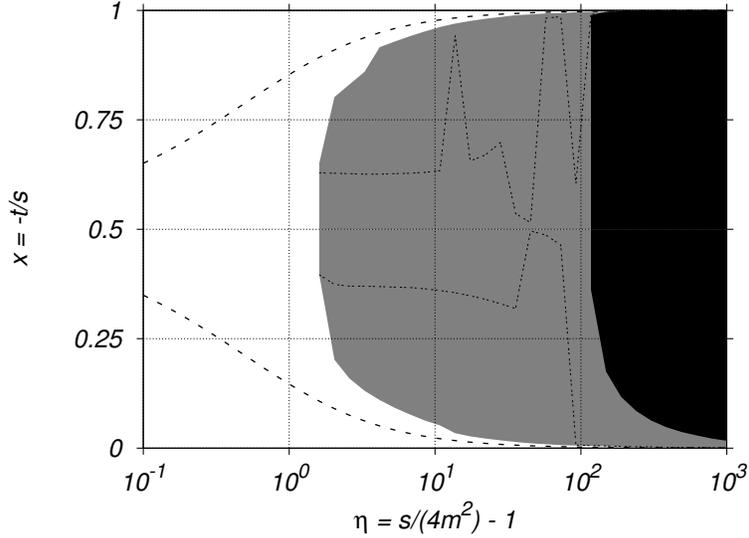,width=10cm}
  \end{center}
  \caption{\label{fig:convergence} Convergence regions of the small mass
    expansion of the two-loop amplitude. The grey area represents the region
    where the series is accurate to one permille, whereas the black area the
    region where the leading term of the series is accurate to one
    percent. The dashed lines delimit the kinematically allowed region,
    whereas the short dashed lines inside the grey area would correspond to
    the convergence region derived according to Eq.~\ref{eq:conditions} without
    the last condition.}
\end{figure}

In order to study the region of convergence and usability of the series, it is
necessary to specify some criteria that could be applied without reference to
any external result. In fact, if the amplitude were approximated with one
permille accuracy, it would be sufficient for any foreseeable
applications. Customarily, the error of an expansion is estimated by the size
of the last term. For geometric series, this estimate is only correct (not
underestimated) if the ratio of two subsequent terms does not exceed $1/2$. In
the latter case, eleven terms of the series (as in the case of the present
result for the amplitude) provide indeed an approximation exact to one
permille. In the case of amplitudes, the series is obviously not geometric,
but conditions inspired by these arguments can be imposed. Let the amplitude
be written as
\begin{equation}
  {\cal A}^{(0,2)} = \sum_{i=0}^{10} a_i(m_s,x) \, m_s^i .
\end{equation}
A relatively conservative heuristic test for the one permille convergence of
the result is given by the following conditions
\begin{equation}
  \label{eq:conditions}
  \left( \left| \frac{a_{10} m_s^{10}}{{\cal A}^{(0,2)}} \right| < 10^{-3} ~
    \bigwedge ~ \left| \frac{a_{10}}{a_9} \right| < \frac{1}{2} ~ \bigwedge ~
  \left| \frac{a_9}{a_8} \right| < \frac{1}{2} \right) ~ \bigvee ~ \left|
  \frac{a_{10} \, m_s^{10}}{{\cal A}^{(0,2)}} \right| < 10^{-5} ,
\end{equation}
which are applied at a given ($m_s$, $x$) point during a scan starting from
the median $x=1/2$, which has always the best convergence for a given value of
$m_s$. The last condition in Eq.~\ref{eq:conditions} deserves further
explanation. It turns out that for relatively small values of $m_s$, the
logarithmic terms in $m_s$ lead to slight violations of the remaining
relations, but the last term of the series is still tiny. In this case, it is
highly improbable that the sum of the missing terms would amount to more than
a permille correction. Without the last test the region of permille
convergence is, therefore, unrealistically restricted. The region resulting
from the application of Eq.~\ref{eq:conditions} is shown in
Fig.~\ref{fig:convergence}. The visible discontinuities of the boundary are
due precisely to the nature of the criterion (and to a lesser extent to the
discretized scan). This figure shows also the region, where the leading term
of the series agrees with the full result to one percent. No special criteria
are needed here, of course, since eleven terms of the series expansion provide
a sufficient approximation to the exact result for the purpose of determining
this region.

\section{Numerical solution}

Since the series expansion does not satisfactorily approximate the result over
the whole range of variation of kinematic parameters, it is only natural to
try to solve the differential equations for the master integrals numerically
and thus obtain a purely numerical description of the amplitude. This idea has
originally been put forward in \cite{Caffo:1998du} for master integrals
corresponding to the two-loop sunrise graph without, however, relating to a
concrete physical problem. In fact, to the best of my knowledge, the only
applications to physical  processes have been attempted in
\cite{Boughezal:2007ny,Czakon:2007qi}. The problem at hand pushes the
difficulty level substantially further, and requires, therefore, a careful
assessment of feasibility. Before I show that high precision numerical results
may indeed be obtained at all relevant points of phase space, there are
several issues that have to be clarified, as it can hardly be overstressed
that the right choice of numerical algorithms will make the difference between
success and failure.

\subsection{Implementation}

At first, it is necessary to determine the type of differential equation
system under consideration. It turns out that different methods are available
for stiff and non-stiff problems. The distinction between the two is somewhat
fuzzy in the professional literature, but there is agreement that stiff
problems involve exponentially decaying solutions. The criterion used in
practice is the existence of large negative eigenvalues of the Jacobian. In
the present case it is, however, easier to use heuristic arguments instead of
performing a numerical analysis. In fact, experience accumulated in numerous
higher order calculations shows that exponentially decaying components would
be rather unusual. I will, therefore, assume without further consideration
that the system is non-stiff. In this case, there is still a large number of
algorithms available. However, because the solutions, {\it i.e.} the master
integrals, must be very smooth (we remain above all thresholds) and high
precision will be requested, a variable coefficient multistep method
\cite{Gear} is expected to be most efficient \cite{NumericalRecipes}. Indeed,
this kind of methods is based on polynomial interpolation/extrapolation with
polynomials of order up to 12, which is a guarantee of very fast convergence
under the assumption of suitable smoothness (if higher order derivatives were
large, the errors would obviously grow severely with the order).

The next choice concerns the treatment of master integrals, as they can be
considered to be either complex or two-component real functions (after
decomposition into real and imaginary parts). Clearly, working with complex
functions reduces the size of the Jacobian, which may give substantial
improvements in the running time. Furthermore, the real function approach
suffers from the fact that the imaginary parts of many integrals vanish for
real arguments, which poses problems as far as error control is concerned.
Indeed, in such cases it is impossible to use a relative error criterion and
absolute errors must be used. The size of the latter can only be determined in
conjunction with the final amplitude in order to have control over the
precision of the outcome. There seems to be only one, albeit very strong,
argument in favor of real functions, namely that most of the available
software works with real numbers exclusively. A quick glance at the literature
of the subject shows, that writing a code from scratch should be
avoided. Fortunately, one of the most advanced software packages implementing
the variable coefficient multistep method \cite{VODE} has recently been
translated to complex arguments.

Closely connected to the choice made above is the problem of error
control. Customarily, working with complex functions implies that the error is
given by the modulus of the difference between the exact value and the
approximation. In the present case, however, we are only interested in the
real part of the amplitude, hence the imaginary components of the master
integrals will be discarded. In consequence, we need to control the error of
the real part and not that of the modulus. Fortunately, unless the imaginary
part is much larger than the real, the two are not much different. It turns
out that in the present calculation, only about 6\% of the evaluated phase
space points involved an integral, for which the imaginary part was more than
$10^4$ larger than the real part. Therefore, for simplicity reasons, the
traditional error estimate has been used in the following. Notice also that
the imaginary parts could have been discarded from the start, since the system
of differential equations is linear and we do not cross any singularities.

After settling the implementation questions, it is necessary to decide on the
position of the boundary conditions. In the original publication
\cite{Caffo:1998du}, it has been proposed to start from a threshold or a
pseudo-threshold, since the values of the integrals at these points were
known. This approach has the drawback that these points are at the same time
singularities of the differential equations, which requires slight
modifications of the algorithm and leads invariably to a substantial loss of
precision when evolving further from the boundary. Here, I use the series
expansion of the previous section to compute the values of the integrals to
very high precision. In fact at
\begin{equation}
  \label{eq:boundary}
  m_s  = 5 \times 10^{-3}, ~~~~ x = \frac{1}{4},
\end{equation}
the relative error estimated by the size of the last term (very conservative)
does never exceed $10^{-18}$. The second condition in Eq.~\ref{eq:boundary}
deserves explanation, because the median point $x = 1/2$ would have led to
better convergence. Unfortunately, we will see later that it is also a
singular point of the system and thus cannot be used. The choice taken results
in the loss of about two digits and is compensated by a twice smaller value of
$m_s$.

\begin{figure}[t]
  \begin{center}
    \epsfig{file=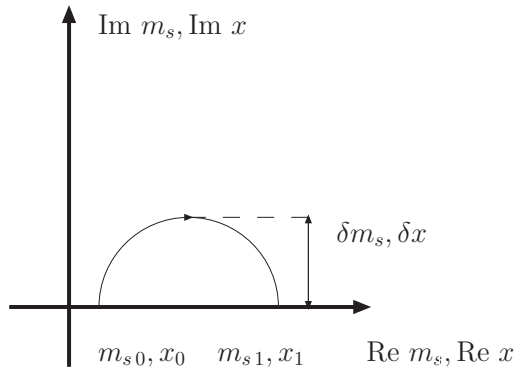,width=7cm}
  \end{center}
  \caption{\label{fig:contours} Evolution contours for the numerical integration
    of the system of differential equations in $m_s$ and $x$.}
\end{figure}

Let me now turn to the discussion of the contour of integration. It is clear
\cite{Caffo:1998du} that evolving along the real axis should be avoided, in
order not to stumble on the singularities. Since we are working with complex
functions anyway, it is not a problem to deform the contour into the complex
plane. In general, this deformation is only restricted by
causality\footnote{for example, if the evolution were performed in the
Mandelstam $s$ variable, the contour should be in the upper half-plane, when
approaching the cut.}. In the present case, however, we always remain above
thresholds, which means that the Riemann sheet has already been chosen when
fixing the boundary conditions, and any curve will be appropriate. To take
full advantage of the multistep algorithm for integration, which depends on
several previous values of the system, it would be desirable to use a single
smooth curve to reach a given point starting from the boundary. Unfortunately,
we need an integration in two variables, and experimentation has shown that it
is more efficient to perform two separate evolutions. For each of these, I
will use an elliptic contour, with a user specified eccentricity as depicted
in Fig.~\ref{fig:contours}. The latter freedom allows for a relatively easy
estimate of the final global error, by computing the desired amplitude with
several different contours. It is also interesting to note that in practice
the convergence turns out to be faster for more circular contours.

\begin{table}[t]
  \begin{center}
    \begin{tabular}{||l|c|c|l||}
      \hline \hline
      Jacobian singularity & branching & allowed & interpretation \\
      \hline \hline
      $m_s = 0$ & yes & & collinear singularity \\
      $m_s = 1/4$ & yes & & s-channel threshold \\
      $m_s = -1/4$ & & & \\
      \hline
      $x = 0$ & yes & & t-channel threshold \\
      $x = 1$ & yes & & u-channel threshold \\
      $x = 1/2$ & & yes & perpendicular scattering \\
      \hline
      $m_s = x \, (1-x)$ & & & forward/backward scattering \\
      $m_s = x$ & & & \\
      $m_s = 1-x$ & & & \\
      $m_s = -x$ & & & \\
      $m_s = x-1$ & & & \\
      \hline
      $m_s = 1/2 \, x \, (1-x)$ & & yes & \\
      $m_s = 1/2 \, x$ & & yes & \\
      $m_s = 1/2 \, (1-x)$ & & yes & \\
      \hline
      $m_s = 1/2 \, (1-x^2)$ & & & \\
      $m_s = -1/2 \, (1-x)^2$ & & & \\
      \hline \hline
    \end{tabular}
  \end{center}
  \caption{\label{tab:singularities} Complete list of singularities of the
    Jacobians, $J^M$ and $J^X$, of the system of differential equations. The
    singularities occur in both systems of differential equations in $m_s$ and
    $x$ apart from the point $m_s = -1/4$, which is present only in the
    differential equations in $m_s$. The table indicates in addition the
    presence of a branching point at a given singularity (a blank entry denotes
    a regular point of the solution), and specifies
    whether a singularity occurs within the kinematically allowed region
    (a blank entry denotes a point outside the allowed region).}
\end{table}

A rather unpleasant feature of the system of differential equations at hand is
the presence of a number of singularities in the Jacobians. They are summarized
in Tab.~\ref{tab:singularities}. It is crucial that aside from thresholds the
solution is regular at these points. Therefore, in the case of the four different
singularities, which occur inside the kinematically allowed region it is
sufficient to resort to interpolation. The problem is more acute for the
singularities at the boundary (forward/backward scattering). These approach
the true branching points at the thresholds when $m \rightarrow 0$, and
therefore interpolation is only efficient for moderate values of $m$, when the
necessary points outside the kinematically allowed region are not trapped too
close to the branching points. Otherwise, extrapolation is necessary. Notice
that the concept of ``dangerous distance to a singularity'' requires
specification. In fact, it is defined by the available numerical precision and
the strength of the singularity (power of the singular polynomial in the
denominator of a coefficient). Finally, one has to remember that the solution
for the master integrals will be input into the expression for the amplitude,
which may lead to further cancellations, and hence instabilities. For the
present calculation, I have adopted extended precision (quadruple) in the
numerics to overcome this problem.

\begin{table}[t]
  \begin{center}
    \begin{tabular}{||l||r|r|r||r|r|r||}
    \hline \hline
    & \multicolumn{3}{c||}{leading color} & \multicolumn{3}{c||}{full color} \\
    \hline \hline 
    number of masters & \multicolumn{3}{c||}{36} & \multicolumn{3}{c||}{145} \\
    number of functions & \multicolumn{3}{c||}{155} & \multicolumn{3}{c||}{595} \\
    \hline \hline
    precision & \multicolumn{2}{c|}{quadruple} & double & \multicolumn{2}{c|}{quadruple} & double \\
    \hline
    \multicolumn{7}{||c||}{evolution in $m_s$} \\
    \hline
    requested local error & $10^{-20}$ & $10^{-12}$ & $10^{-12}$ & $10^{-20}$ & $10^{-12}$ & $10^{-12}$ \\
    contour deformation $\delta m_s$ & 0.1 & 0.1 & 0.1 & 0.1 & 0.1 & 0.1 \\
    number of steps taken & 2319 & 618 & 534 & 2932 & 777 & 1302 \\
    Jacobian evaluation time [ms] & 3.4 & 3.4 & 0.2 & 37 & 37 & 4.9 \\
    \hline
    \multicolumn{7}{||c||}{evolution in $x$} \\
    \hline
    requested local error & $10^{-18}$ & $10^{-10}$ & $10^{-10}$ & $10^{-18}$ & $10^{-10}$ & $10^{-10}$ \\
    contour deformation $\delta x$ & 0.1 & 0.1 & 0.1 & 0.1 & 0.1 & 0.1 \\
    number of steps taken & 545 & 139 & 139 & 739 & 174 & 432 \\
    Jacobian evaluation time [ms] & 8.3 & 8.3 & 0.4 & 150 & 150 & 17 \\
    \hline \hline
    total evaluation time [s]& 49 & 13 & $<$ 1 & 957 & 243 & 26 \\
    \hline \hline
    \end{tabular}
    \caption{\label{tab:efficiency} Timing and efficiency information for the
      numerical integration of the system of differential equations to the
      point $m_s = .2$, $x = .45$. The numbers have been obtained on a 2GHz
      Intel Core 2 Duo system, after compilation with the Intel Fortran
      compiler. Quadruple precision is an option of the compiler.}
  \end{center}
\end{table}

\begin{table}[t]
  \begin{center}
    \begin{tabular}{||l|r|r|r|r|r||}
      \hline \hline
      & $\ep^{-4}$ & $\ep^{-3}$ & $\ep^{-2}$ & $\ep^{-1}$ & $\ep^{0}$ \\
      \hline \hline
      $A$ & 0.22625 & 1.391733154 & -2.298174307 & -4.145752449 & 17.37136599 \\
      $B$ & -0.4525 & -1.323646320 & 8.507455541 & 6.035611156 & -35.12861106 \\
      $C$ & 0.22625 & -0.06808683395 & -18.00716652 & 6.302454931 & 3.524044913 \\
      $D_l$ &  & -0.22625 & 0.2605057339 & -0.7250180282 & -1.935417247 \\
      $D_h$ &  &  & 0.5623350684 & 0.1045606449 & -1.704747998 \\
      $E_l$ &  & 0.22625 & -0.3323207300 & 7.904121951 & 2.848697837 \\
      $E_h$ &  &  & -0.5623350684 & 4.528240788 & 12.73232424 \\
      $F_l$ &  &  &  &  & -1.984228442 \\
      $F_{lh}$ &  &  &  &  & -2.442562819 \\
      $F_h$ &  &  &  &  & -0.07924540546 \\
      \hline \hline
    \end{tabular}
    \caption{\label{tab:values} Values of the color coefficients of the
      two-loop amplitude at the point $m_s = .2$, $x = .45$ rounded at
      10 digits precision (the given digits are unaffected by numerical
      uncertainties).}
  \end{center}
\end{table}

\subsection{Efficiency and numerical stability}

In order to illustrate the efficiency of the approach, I show in
Tab.~\ref{tab:efficiency} different timings and other related informations for
the complete solution at the point $m_s = .2$ and $x = .45$. Since the
evolution is performed separately first in $m_s$ and then in $x$, I require
two more digits of local precision in the first step, so that the estimate of
the error will be dominated by the second. Note that in the first evolution
the $x$ value is fixed from the very beginning as specified in the boundary
condition Eq.~\ref{eq:boundary}. This value has been hardcoded in the
Jacobian, which not only leads to a much faster evaluation, but also to less
severe numerical instabilities in the coefficients themselves. The latter
feature is actually crucial in reaching the higher precision in the first
evolution. Returning to the error estimate, it has to be stressed that in
numerical integration of systems of differential equations, the precision is
specified locally, {\it i.e.} one requires that the error of the approximation
does not exceed a certain value at every step. Therefore, the global relative
error is estimated by the product of the number of steps taken and the
requested local error. In practice, the number of steps in the $m_s$ evolution
never exceeded ten thousand, and therefore the final precision for a local
relative error of $10^{-20}$ (in quadruple precision) was roughly sixteen
digits. The number of steps needed in the second evolution is usually smaller,
because the starting point is far from any singularities. In consequence for
the requested local error of $10^{-18}$, the final global error should not
exceed about $10^{-15}$. There is one additional source of error connected to
roundoff and numerical cancellations in the coefficients. Its presence is
visible in the table, when comparing the solution in double precision and
quadruple precision for the same requested local relative error. In the case
of the $x$ evolution of the full system, the number of steps is larger in
double precision, precisely because the error estimates are not satisfied due
to random roundoff errors. The software tries to reduce the step size until
the error estimate satisfies the bound, which eventually happens because the
random variations around the true value must, sooner or later, turn close to
it. Let me finally comment on the running time. Clearly, if only the leading
color coefficient were needed with moderate precision, a value at a single
phase space point could be obtained within less than a second. For the full
color structure in quadruple precision, as much as fifteen minutes are
needed. Although this does not allow for direct implementation in a
Monte-Carlo generator, the functions are smooth enough to be interpolated
starting from a grid of values. The efficiency is by far sufficient to obtain
dense grids. Therefore, a grid of numerical values for moderate $m_s$ together
with the series expansion of the previous section for small $m_s$ is a
complete solution to the problem of evaluation of the two-loop amplitudes for
the production of a heavy quark-anti-quark pair in massless quark-anti-quark
annihilation. In Tab.~\ref{tab:values}, I give the values of all the color
coefficients with ten digits precision at the point $m_s = .2$, $x = .45$,
which is well outside the region of convergence of the series expansion. The
actual precision of the result estimated by the variation with respect to the
change of the integration contour in $x$ was roughly fourteen digits.

\begin{figure}[t]
  \begin{center}
    \epsfig{file=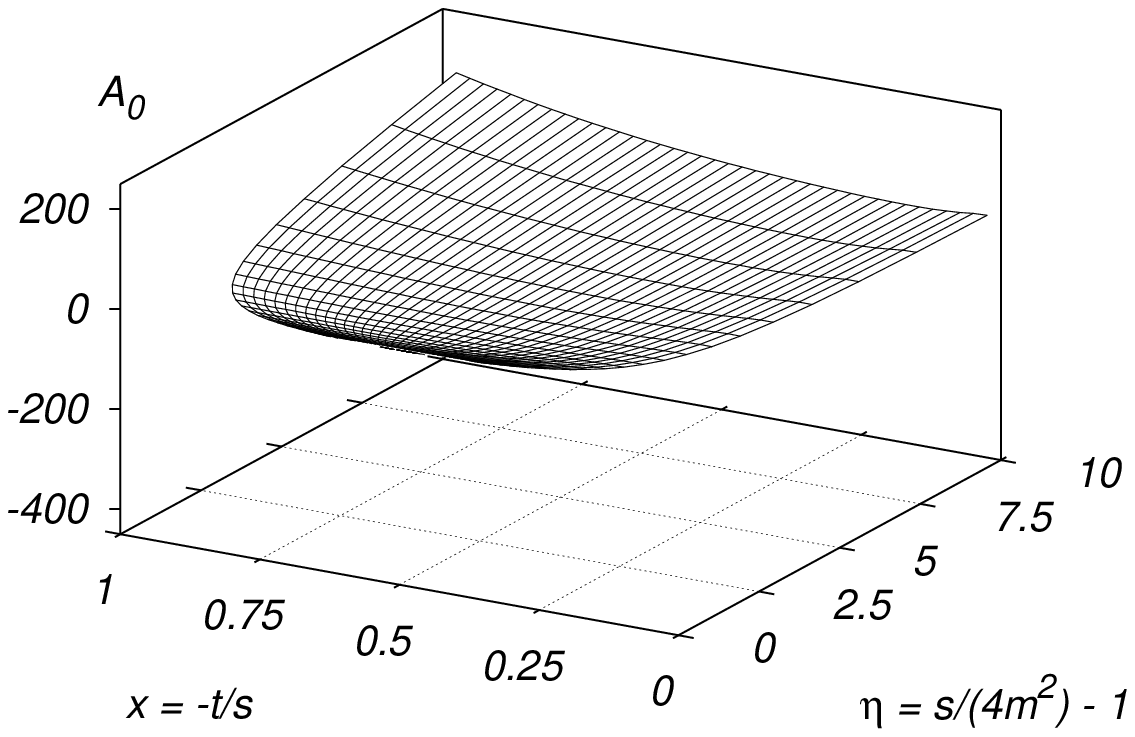,width=8cm}\epsfig{file=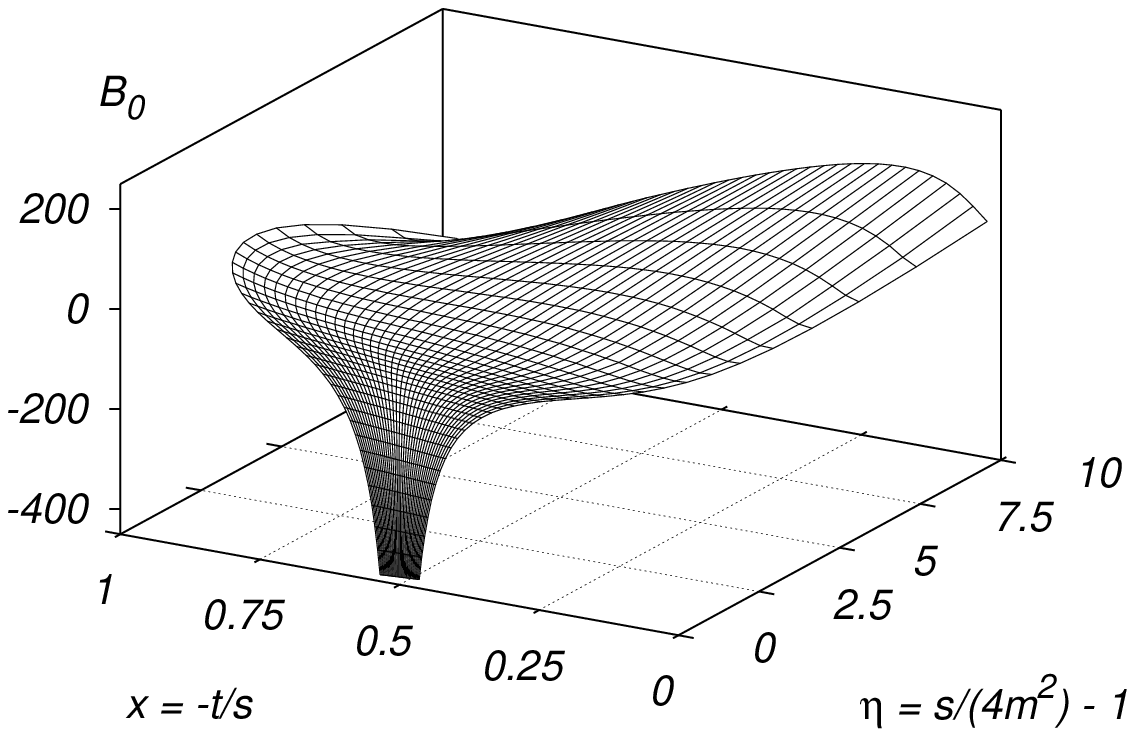,width=8cm}
    \epsfig{file=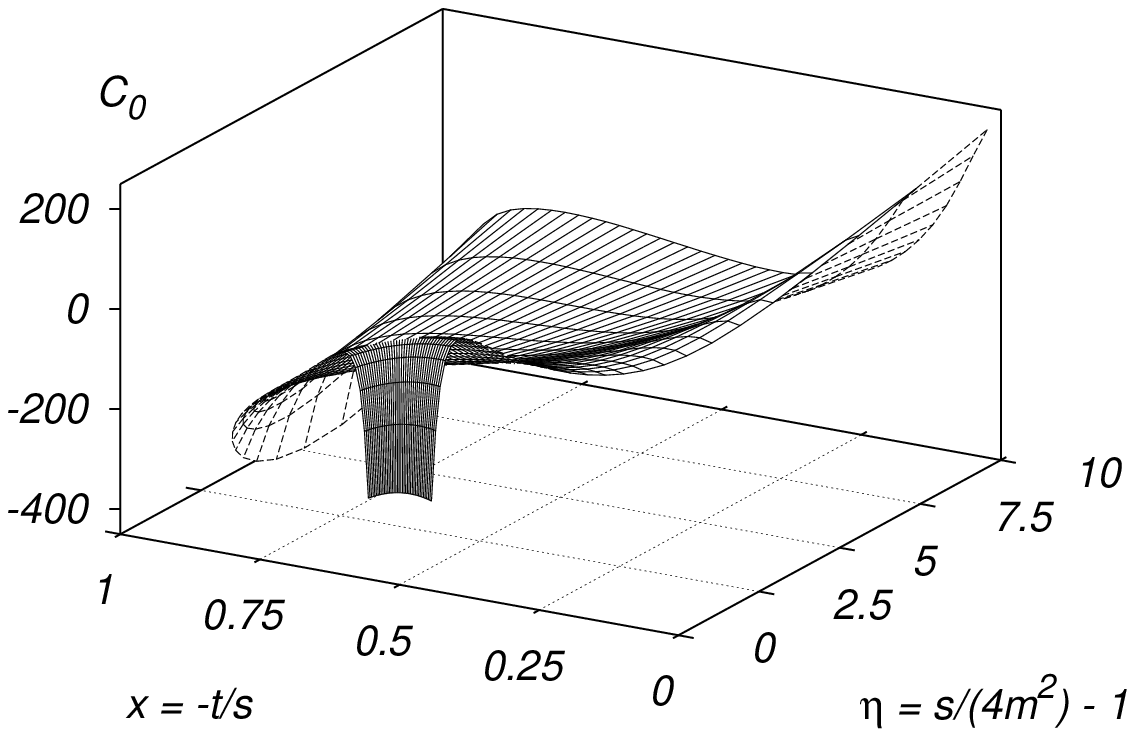,width=8cm}\epsfig{file=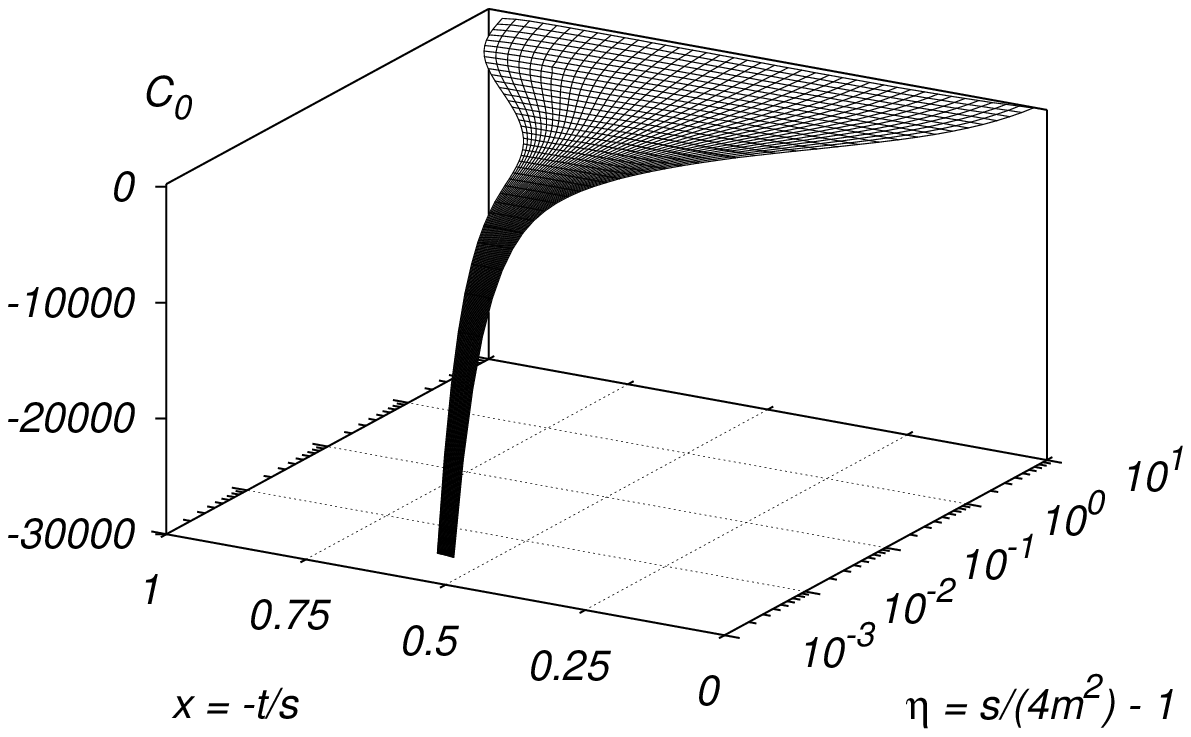,width=8cm}
  \end{center}
  \caption{\label{fig:plots1} Finite parts of the bosonic contributions to the
  two-loop amplitude.}
\end{figure}

\begin{figure}[t]
  \begin{center}
    \epsfig{file=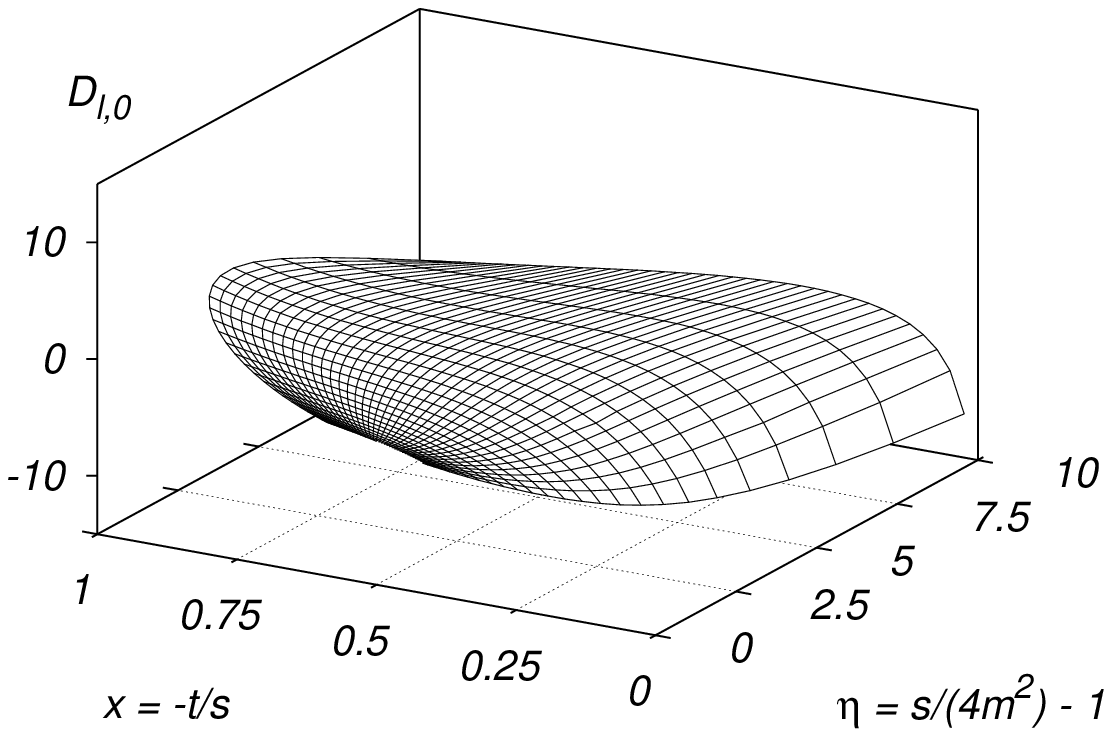,width=8cm}\epsfig{file=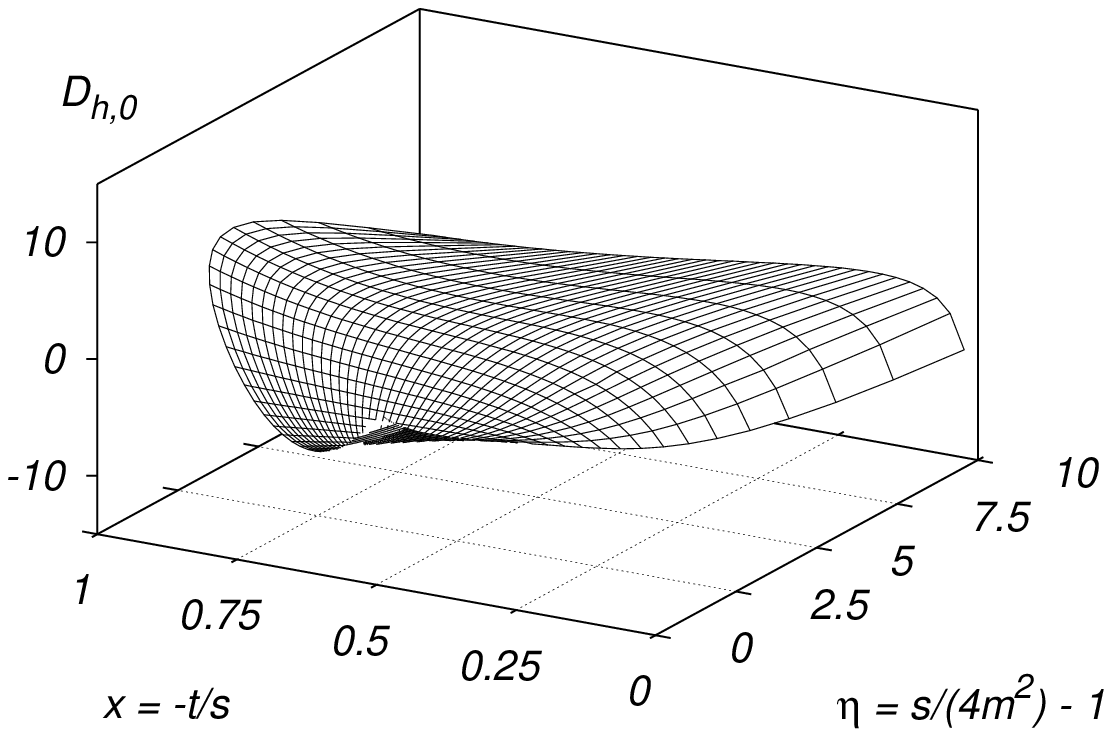,width=8cm}
    \epsfig{file=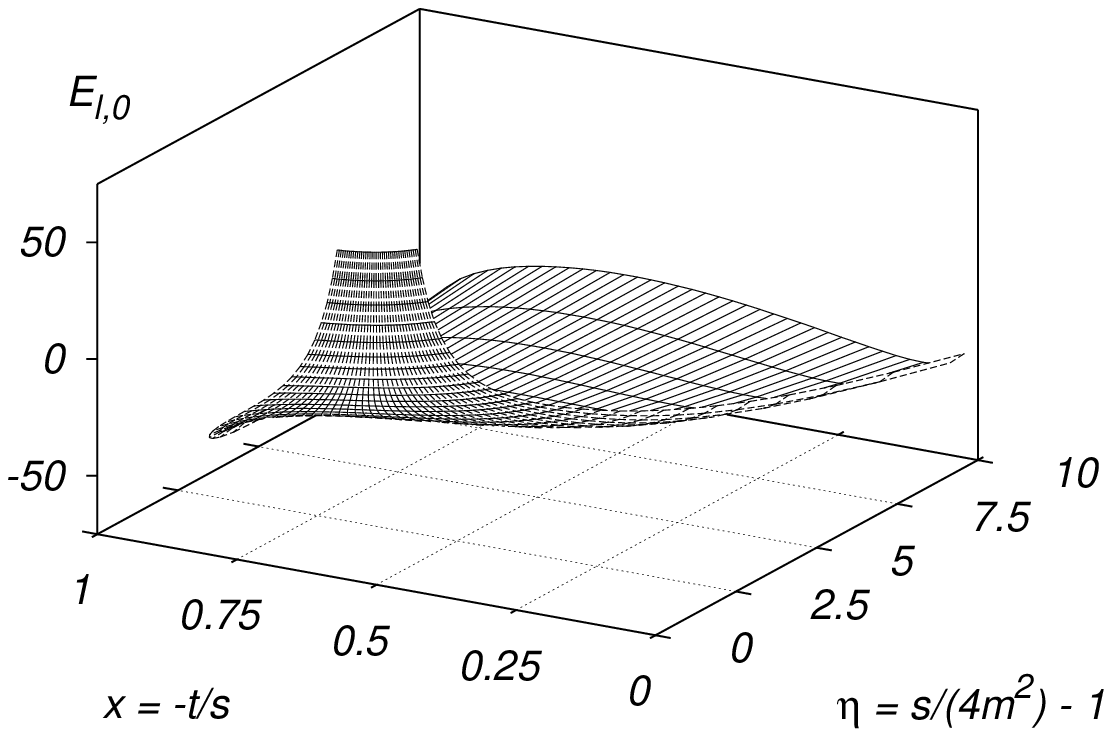,width=8cm}\epsfig{file=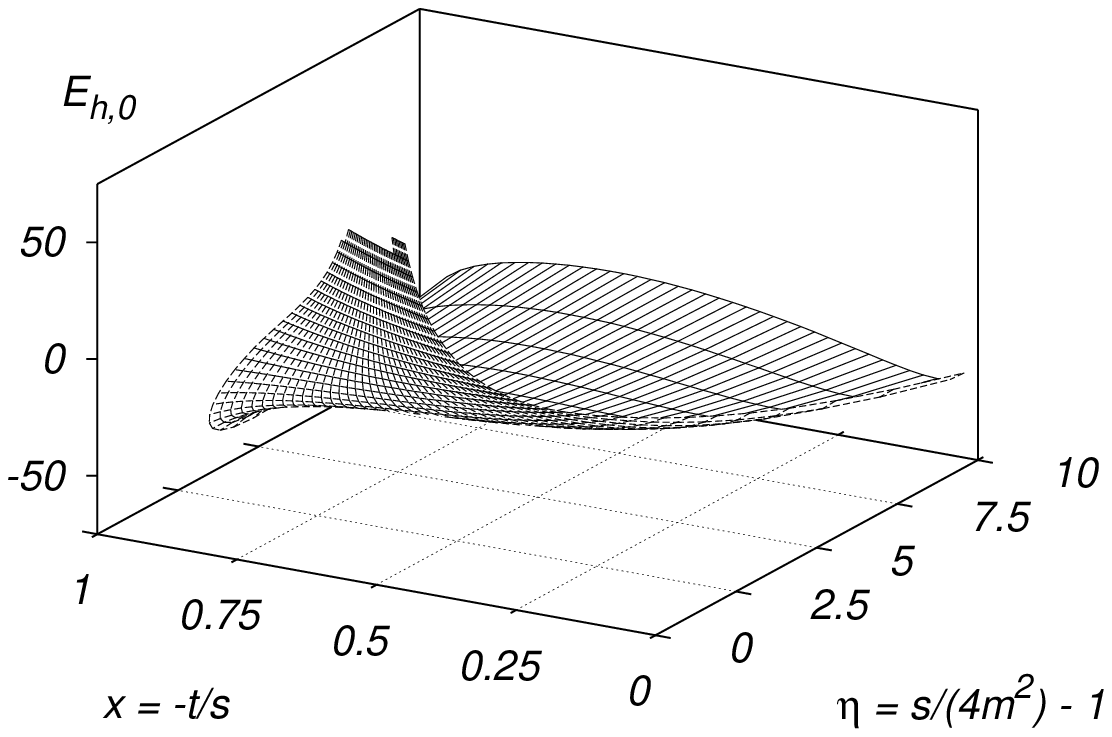,width=8cm}
  \end{center}
  \caption{\label{fig:plots2} Finite parts of the single fermionic
    contributions to the two-loop amplitude.}
\end{figure}

In the course of preparation of the present work, I derived the necessary
grids mentioned above. For simplicity the singularities at the kinematic
boundaries (forward/backward scattering) have been avoided by keeping a
distance of $10^{-3}$ in $x$, which is unnoticeable on the plots, but can be
improved upon in the future. In this respect, the actual needs will only be
apparent once the virtual corrections will be combined with real
radiation. The range of variation in $m_s$ was chosen such that the distance
to the threshold parameterized by
\begin{equation}
\eta = \frac{s}{4m^2}-1
\end{equation}
was at least $10^{-3}$. This is safely sufficient for any practical
applications. The plots of the finite parts of the purely bosonic corrections
can be found in Fig.~\ref{fig:plots1}, whereas those for the contributions
containing a single closed fermionic loop in Fig.~\ref{fig:plots2}. The
interesting feature is the large variation of the subleading color
contributions, when nearing the threshold. Of course, this is due to the
Coulomb singularity. In particular, the $C_0$ coefficient behaves like
$1/\beta^2$, which cannot be compensated by phase space integration. This
leads to a true divergence, to be taken care of by resummation in the complete
analysis.

Finally, let me point two possible improvements of the implementation. The
first one is related to the numerical precision. Clearly, different choices of
the basis functions may be used to milden the strength of the
singularities. For example, for two functions, which suffer from large
cancellations, one could introduce a mixture such that one of the functions is
small, while the other is large, but contributes with a small
coefficient. The second possible improvement is at a lower level and concerns
the time of evaluation. Since the computation is done in quadruple precision,
one could try different libraries instead of the compiler's built-in
routines. In fact, the implementation of \cite{QD} has proven to be about three
times faster on some problems. Of course, this is of lesser importance than
precision, since even times of the order of five minutes per point would
still not allow to perform the integration in real time within the framework
of a Monte Carlo generator.

\section{Conclusions}

In this paper, I have presented a complete solution to the problem of
evaluation of the two-loop amplitude for heavy quark production in light quark
annihilation. The result, a numeric approximation obtained by a combination of
a small mass expansion and integration of a system of differential equations,
is satisfactory from the point of view of applications. Aiming at a complete
description of the top quark pair production cross section at the
next-to-next-to-leading order at the LHC, the next steps to perform will be
the evaluation of the two-loop amplitude in gluon fusion and the computation
of the real radiation contributions. The mixed real-virtual contributions are
in principle known, but have to be supplied with suitable subtraction terms in
order to allow for integration over the full phase space.

The result for the series expansion of the amplitude, as well as the grid of
values obtained by numerical integration of differential equations (with all
numbers rounded at 5 digits) are available in the form of {\sc Mathematica}
files, {\tt qqQQ2Lseries.m} and {\tt qqQQ2Lnumeric.m} respectively, attached to
the source of the paper on the preprint server {\tt http://arXiv.org}, but can
also be obtained from the author upon request.

\section*{Acknowledgments}

This work was supported by the Sofja Kovalevskaja Award of the Alexander von
Humboldt Foundation.


\end{document}